\documentclass[twocolumn,amsmath,amssymb,superscriptaddress,showkeys]{revtex4}

\usepackage{verbatim} 

\usepackage{graphicx}% Include figure files
\usepackage{amstext,amsmath,amssymb,amsfonts}
\usepackage{bm}% bold math

\usepackage{txfonts}
\usepackage{mathtools}
\usepackage{hyperref}
\usepackage{color}
\definecolor{lightblue}{rgb}{0.2,0.2,0.7}
\definecolor{darkblue}{rgb}{0,0.25,0.5}
\definecolor{redbrown}{rgb}{0.875,0.25,0.125}
\definecolor{darkgreen}{rgb}{0,0.5,0}

\renewcommand{\b}[1]{\ensuremath{\mathbf{#1}}}

\newcommand{\sr}{\ensuremath{\text{sr}}}

\newcommand{\HF}{\ensuremath{\text{HF}}}

\renewcommand{\d}{\ensuremath{\text{d}}}

\renewcommand{\c}{\ensuremath{\text{c}}}

\newcommand{\md}{\ensuremath{\text{md}}}
\DeclareMathOperator{\erf}{erf}

\newcommand{\B}{\ensuremath{{\cal B}}}

\begin{document}

\title{A density-fitting implementation of the density-based basis-set correction method}

\author{Andreas He{\ss}elmann}
\email{andreas.hesselmann@theochem.uni-stuttgart.de}
\affiliation{Institute for Theoretical Chemistry, University of Stuttgart, 70569, Stuttgart, Germany}

\author{Emmanuel Giner}
\email{emmanuel.giner@lct.jussieu.fr}
\affiliation{Laboratoire de Chimie Théorique, Sorbonne Université and CNRS, F-75005 Paris, France}

\author{Peter Reinhardt}
\email{peter.reinhardt@lct.jussieu.fr}
\affiliation{Laboratoire de Chimie Théorique, Sorbonne Université and CNRS, F-75005 Paris, France}

\author{Peter J. Knowles}
\email{KnowlesPJ@Cardiff.ac.uk}
\affiliation{School of Chemistry, Cardiff University, Cardiff CF10 3AT, United Kingdom}

\author{Hans-Joachim Werner}
\email{werner@theochem.uni-stuttgart.de}
\affiliation{Institute for Theoretical Chemistry, University of Stuttgart, 70569, Stuttgart, Germany}

\author{Julien Toulouse}
\email{toulouse@lct.jussieu.fr}
\affiliation{Laboratoire de Chimie Théorique, Sorbonne Université and CNRS, F-75005 Paris, France}
\affiliation{Institut Universitaire de France, F-75005 Paris, France}

\date{December 11, 2023}

\begin{abstract}
This work reports an efficient density-fitting implementation of the density-based basis-set correction (DBBSC) method in the \textsc{MOLPRO} software. This method consists in correcting the energy calculated by a wave-function method with a given basis set by an adapted basis-set correction density functional incorporating the short-range electron correlation effects missing in the basis set, resulting in an accelerated convergence to the complete-basis-set limit. Different basis-set correction density-functional approximations are explored and the complementary-auxiliary-basis-set single-excitation correction is added. The method is tested on a benchmark set of reaction energies at the second-order M{\o}ller-Plesset (MP2) level and a comparison with the explicitly correlated MP2-F12 method is provided. The results show that the DBBSC method greatly accelerates the basis convergence of MP2 reaction energies, without reaching the accuracy of the MP2-F12 method but with a lower computational cost.
\end{abstract}


\maketitle

\section{Introduction}

One of the main goals of quantum chemistry is the accurate prediction of molecular properties, which requires to tackle the electron correlation problem. For this, there are two main families of computational electronic-structure methods: wave-function theory (WFT)~\cite{Pop-RMP-99} which targets the complicated many-electron wave function, and density-functional theory (DFT)~\cite{Koh-RMP-99} which uses the simpler one-electron density. 
While DFT has become the workhorse of quantum chemistry thanks to its appealing balance 
between computational cost and accuracy, the lack of a systematic scheme to improve the quality of 
density-functional approximations has renewed the interest in the development of WFT methods in the last few decades.

A serious limitation of WFT methods is their slow convergence of the correlation energy with the size of the one-electron basis set. This slow convergence originates from the short-range singularity of the Coulomb electron-electron repulsion which induces a derivative discontinuity in the exact eigenstate wave functions, known as the electron-electron cusp condition~\cite{Kat-CPAM-57}. There are two main approaches for dealing with this problem. The first approach consists in extrapolating the results to the complete-basis-set (CBS) limit by using increasingly large basis sets~\cite{HelKloKocNog-JCP-97,HalHelJorKloKocOlsWil-CPL-98}. The second approach consists in using explicitly correlated R12 or F12 methods which incorporate in the wave function a correlation factor reproducing the electron-electron cusp (see, e.g., Refs.~\onlinecite{KloSam-JCP-02,Man-JCP-03,Ten-JCP-04,Ten-CPL-04,Val-CPL-04,WerAdlMan-JCP-07,KniAdlWer-JCP-09,TenNog-WIRES-12,HatKloKohTew-CR-12,KonBisVal-CR-12,ShiWer-MP-13}).

An alternative approach to accelerate basis-set convergence was recently proposed, which we will refer as the density-based basis-set correction (DBBSC) method~\cite{GinPraFerAssSavTou-JCP-18}. It consists in correcting the energy calculated by a WFT method with a given basis set by an adapted basis-set correction density functional incorporating the short-range electron correlation effects missing in the basis set, resulting in an accelerated convergence to the CBS limit. In practice, this basis-set correction density functional is constructed from range-separated DFT~\cite{TouColSav-PRA-04} by defining a basis-dependent local range-separation parameter which provides a local measure of the incompleteness of the basis set. This DBBSC method was validated for configuration-interaction and coupled-cluster calculations of atomization energies~\cite{LooPraSceTouGin-JPCL-19,YaoGinLiTouUmr-JCP-20,YaoGinAndTouUmr-JCP-21}, excitation energies~\cite{GinSceTouLoo-JCP-19}, dissociation energy curves~\cite{GinSceLooTou-JCP-20}, and dipole moments~\cite{GinTraPraTou-JCP-21,TraTouGin-JCP-22}. It was also extended to GW calculations~\cite{LooPraSceGinTou-JCTC-20} and to linear-response theory~\cite{TraGinTou-JCP-23}, and some mathematical aspects of the method were studied in detail on a one-dimensional model system~\cite{TraGinTou-JCP-22}.

In this work, we report an efficient implementation of the DBBSC method in the \textsc{MOLPRO} software~\cite{WerKnoKniManSch-WIR-12,WerKnoManBlaDolHesKatKohKorKreMaMilMitPetPolRauSib-JCP-20,Molproshort-PROG-23} in which density fitting~\cite{WerManKno-JCP-03} is used to alleviate the computational bottleneck of the method, namely the calculation of the local range-separation parameter. This allows us to use the DBBSC method on larger molecular systems than what was previously possible. We thus apply the DBBSC method for correcting the basis-set errors in the molecular reaction energies of the FH51 benchmark set~\cite{FriHan-JCTC-13,Fri-JCTC-15} at the second-order M{\o}ller-Plesset (MP2) level. We also test different basis-set correction density-functional approximations, as well as the addition of a single-excitation correction for one-electron basis-set errors. Finally, we compare the performance of the DBBSC method with the explicitly correlated MP2-F12 method~\cite{WerAdlMan-JCP-07}.
 
The paper is organized as follows. In Section~\ref{sec:theory}, we explain the theory of the present implementation of the DBBSC method. Section~\ref{sec:details} provides computational details for the calculations on the FH51 benchmark set. In Section~\ref{sec:results}, we give and discuss our results. Finally, Section~\ref{sec:conclusion} contains our conclusions.

\section{Theory}
\label{sec:theory}

For simplicity, we give the equations for closed-shell states and we assume real-valued HF spatial orbitals $\{\varphi_p \}$.

\subsection{The DBBSC method at the MP2 level}

Given the MP2 total energy $E_\text{MP2}^\B$ in a basis set $\B$, we apply the non-self-consistent basis-set correction~\cite{GinPraFerAssSavTou-JCP-18,LooPraSceTouGin-JPCL-19} as
\begin{equation}
E_\text{MP2+DFT}^\B = E_\text{MP2}^\B + \bar{E}^\B[n_\HF^\B],
\label{EMP2+DFT}
\end{equation}
where $\bar{E}^\B[n_\HF^\B]$ is the basis-correction density functional evaluated at the active HF density $n_\HF^\B$ (i.e., excluding core orbitals in case of frozen-core calculations). In order not to affect the CBS limit, this functional $\bar{E}^\B[n]$ must be such that it vanishes when the basis set $\B$ is complete. Moreover, provided a good enough approximation is used for $\bar{E}^\B[n]$, the basis-set corrected MP2 energy, referred to as ``MP2+DFT'', is expected to converge faster to the MP2 CBS limit than the uncorrected MP2 energy.

\subsection{Local range-separation parameter}

The dependence on the basis set of the basis-correction density functional $\bar{E}^\B[n]$ comes from the local range-separation parameter $\mu^\B(\b{r})$. It is defined as~\cite{GinPraFerAssSavTou-JCP-18,LooPraSceTouGin-JPCL-19}
\begin{equation}
\mu^\B(\b{r}) = \frac{\sqrt{\pi}}{2} W^\B(\b{r}),
\end{equation}
where $W^\B(\b{r})$ is the on-top value of the effective interaction localized with the HF wave function
\begin{equation}
W^\B(\b{r}) = 
\begin{cases}
\frac{f^\B_\HF(\b{r})}{n_{2,\HF}^\B(\b{r})}, & \text{if} \; n_{2,\HF}^\B(\b{r}) \neq 0,\\
\infty, & \text{otherwise}.
\end{cases}
\label{WBr}
\end{equation}
In Eq.~(\ref{WBr}), $n_{2,\HF}^\B(\b{r})$ is the HF on-top pair density
\begin{equation}
n_{2,\HF}^\B(\b{r}) = \frac{n_\HF^\B(\b{r})^2}{2},
\end{equation}
with the active HF density $n_\HF^\B(\b{r}) = 2 \sum_{i}^\text{act}\varphi_{i}(\b{r})^2$, and $f^\B_\HF(\b{r})$ has the expression
\begin{equation}
f^\B_\HF(\b{r}) = 2 \sum_{p,q}^\text{all} \sum_{i,j}^\text{act} \varphi_{p}(\b{r}) \varphi_{i}(\b{r}) (\varphi_p \varphi_i | \varphi_q \varphi_j ) \varphi_{q}(\b{r}) \varphi_{j}(\b{r}),
\label{fBHFr}
\end{equation}
where $p$ and $q$ run over all (occupied + virtual) HF spatial orbitals, $i$ and $j$ run over active HF spatial orbitals, and 
$(\varphi_p \varphi_i | \varphi_q \varphi_j )$ are the two-electron Coulomb integrals in chemists' notation. We recall that by active orbitals we mean occupied orbitals without the frozen-core orbitals, in case of frozen-core calculations.

The local range-separation parameter $\mu^\B(\b{r})$ provides a local measure of the incompleteness of the basis set. A straightforward calculation of $f^\B_\HF(\b{r})$ in Eq.~(\ref{fBHFr}) requires to first calculating the molecular-orbital two-electron integrals $(\varphi_p \varphi_i | \varphi_q \varphi_j )$ with a dominant scaling of $O(N_\text{act} N_\text{all}^4)$, and then performing the sums at each grid point which scales as $O(N_\text{act}^2 N_\text{all}^2 N_\text{grid})$, where $N_\text{act}$ is the number of active orbitals, $N_\text{all}$ is the total number of orbitals in the basis, and $N_\text{grid}$ is the number of spatial grid points. This is the computational bottleneck of the basis-set correction calculation.

This scaling can be reduced by density fitting~\cite{Whi-JCP-73,WerManKno-JCP-03}. Introducing an auxiliary fitting basis set $\{ \chi_A \}$, the orbital product is approximated as
\begin{equation}
\varphi_{p}(\b{r}) \varphi_{i}(\b{r}) \approx \sum_{A}^\text{fit} d_A^{pi} \chi_A(\b{r}),
\end{equation}
where $d_A^{pi}$ are the Coulomb-fitting coefficients
\begin{equation}
d_B^{pi} = \sum_{A}^\text{fit} (\varphi_p \varphi_i| \chi_A) [\b{J}^{-1}]_{AB},
\end{equation}
with
\begin{equation}
J_{AB} = \iint \frac{\chi_A(\b{r}_1) \chi_B(\b{r}_2)}{||\b{r}_2 - \b{r}_1||} \d\b{r}_1 \d\b{r}_2,
\end{equation}
and
\begin{equation}
(\varphi_p \varphi_i|\chi_A) = \iint \frac{ \varphi_{p}(\b{r}_1) \varphi_{i}(\b{r}_1)\chi_B(\b{r}_2)}{||\b{r}_2 - \b{r}_1||} \d\b{r}_1 \d\b{r}_2.
\end{equation}
Orthonormalizing the auxiliary fitting basis functions with respect to the metric $\b{J}$,
\begin{equation}
\tilde{\chi}_A = \sum_{B}^\text{fit} [\b{J}^{-1/2}]_{AB} \; \chi_B,
\end{equation}
we can approximate the two-electron integrals as
\begin{equation}
(\varphi_p \varphi_i | \varphi_q \varphi_j) \approx \sum_{A}^\text{fit} (\varphi_p \varphi_i|\tilde{\chi}_A) (\tilde{\chi}_A|\varphi_q \varphi_j),
\label{piqjapprox}
\end{equation}
and the quantity $f^\B_\HF(\b{r})$ in Eq.~(\ref{fBHFr}) as
\begin{equation}
f^\B_\HF(\b{r}) \approx 2 \sum_{A}^\text{fit} \left[ \sum_p^\text{all} \sum_{i}^\text{act} \varphi_{p}(\b{r}) \varphi_{i}(\b{r}) (\varphi_p \varphi_i|\tilde{\chi}_A) \right]^2.
\label{fBHFrapprox}
\end{equation}
Thus, with density fitting, there is no need to build explicitly the two-electron integrals anymore and the calculation of $f^\B_\HF(\b{r})$ in Eq.~(\ref{fBHFrapprox}) now scales as $O(N_\text{act} N_\text{all} N_\text{fit} N_\text{grid})$ where $N_\text{fit}$ is the number of auxiliary fitting basis functions. In practice, the same auxiliary fitting basis sets optimized for density fitting in MP2 can be used here.

\subsection{Approximate basis-correction density functional}
\label{sec:approxfunc}

We approximate the basis-correction density functional with the local form~\cite{LooPraSceTouGin-JPCL-19}
\begin{equation}
\bar{E}^\B[n] \approx \int \bar{e}_{\c,\md}^\sr(n(\b{r}),\nabla n(\b{r}),\mu^\B(\b{r})) \d \b{r},
\label{EBnlocal}
\end{equation}
where $\bar{e}_{\c,\md}^\sr(n,\nabla n,\mu)$ is the complementary multi-determinant short-range correlation functional energy density~\cite{FerGinTou-JCP-19,LooPraSceTouGin-JPCL-19}
\begin{equation}
\bar{e}_{\c,\md}^\sr(n,\nabla n,\mu^\B) = \frac{e_\c(n,\nabla n)}{1+ \frac{e_\c(n,\nabla n)}{c \; n_2(n)}\mu^3},
\label{ecmdsr}
\end{equation}
where $c=(2\sqrt{\pi}(1-\sqrt{2}))/3$ and $n_2(n)$ is a model of the on-top pair density. In Eq.~(\ref{ecmdsr}), $e_\c(n,\nabla n)$ is a standard Kohn-Sham correlation functional energy density. As in previous works, the default choice is the PBE correlation functional~\cite{PerBurErn-PRL-96}. In this work, we also test using the LDA~\cite{PerWan-PRB-92}, LYP~\cite{LeeYanPar-PRB-88}, TPSS~\cite{TaoPerStaScu-PRL-03}, and SCAN~\cite{SunRuzPer-PRL-15} correlation functionals. Note that the TPSS and SCAN functionals are meta-GGA functionals, i.e. they depend also on the non-interacting positive kinetic energy density $\tau(\b{r}) = (1/2) \sum_i^\text{act} |\nabla \varphi(\b{r})|^2$, and thus constitute a slight extension of Eqs.~(\ref{EBnlocal}) and~(\ref{ecmdsr}).

The default choice~\cite{LooPraSceTouGin-JPCL-19} for $n_2(n)$ is to use the on-top pair density of the uniform-electron gas (UEG) 
\begin{equation}
n_2^\text{UEG}(n) = n^2 g_0(n),
\end{equation}
where the on-top pair-distribution function $g_0(n)$ is parametrized in Eq.~(46) of Ref.~\onlinecite{GorSav-PRA-06}. In this work, we also explore two other on-top pair-density models. The first one is the Colle-Salvetti (CS) model~\cite{ColSal-TCA-75,ColSal-TCA-79,MosSanPas-TCA-06}
\begin{equation}
n_2^\text{CS}(n) = \frac{n^2}{2} \Phi_\text{CS}(n)^2,
\end{equation}
where
\begin{equation}
\Phi_\text{CS}(n) = \frac{\sqrt{\pi} \; \beta(n)}{1+\sqrt{\pi} \; \beta(n)},
\end{equation}
and
\begin{equation}
\beta(n) = q \; n^{1/3},
\label{betan}
\end{equation}
where $q$ is an empirical parameter. The second one is the Hollett-Pegoretti (HP) model~\cite{HolPeg-JCP-18}
\begin{equation}
n_2^\text{HP}(n) = \frac{n^2}{2} \Phi_\text{HP}(n),
\end{equation}
where 
\begin{equation}
\Phi_\text{HP}(n) = \frac{2\sqrt{\pi}\;\beta(n)^2}{2 \beta(n) e^{-\frac{1}{4 \beta(n)^2}} + \sqrt{\pi} \left(1+ 2\beta(n)^2\right) \left[1+\erf\left(\frac{1}{2\beta(n)}\right) \right]}.
\end{equation}
We may choose the value of the parameter $q$, e.g., by imposing that the integral of the model on-top pair density equals the integral of the exact on-top pair density, $\int n_2^\text{model}(n(\b{r})) \d\b{r} = \int n_2^\text{exact}(\b{r}) \d\b{r}$, in the helium atom. Estimating $n_2^\text{exact}(\b{r})$ with a highly accurate 418-term Hylleraas-type wave function~\cite{FreHuxMor-PRA-84,BakFreHilMor-PRA-90,UmrGon-PRA-94}, we find $q=1.88$ for the CS model and $q=2.05$ for the HP model. When these on-top pair-density models are used with the PBE correlation functional in Eq.~(\ref{ecmdsr}), we call the resulting basis-set correction functionals PBE-CS and PBE-HP, respectively.

\subsection{CABS single-excitation correction}

For small basis sets $\B$, the HF energy can have a substantial basis-set error. This HF basis-set error is not corrected by the approximate basis-set correction functionals in Section~\ref{sec:approxfunc} since they only correct for missing short-range correlation. The HF basis-set error can however be easily corrected by using the complementary auxiliary basis set (CABS) (see, e.g., Refs.~\cite{Val-CPL-04,AdlKniWer-JCP-07,KniWer-JCP-08,KniAdlWer-JCP-09,BisWolTewKlo-MP-09,NogSim-CP-09,ShaHil-JCTC-17}) used in explicitly correlated R12/F12 methods. In this approach, a large orthonormal basis set is formed by the occupied+virtual HF orbitals obtained in the normal basis set $\B$ and an additional set of virtual orbitals obtained from the CABS. The HF energy correction due to the addition of the CABS is estimated by second-order perturbation theory, leading to the expression, in a closed-shell formalism,~\cite{AdlKniWer-JCP-07,KniWer-JCP-08,KniAdlWer-JCP-09}
\begin{equation}
\Delta E_\text{HF}^{\B,\text{CABS}} = 2 \sum_i^\text{act} \sum_\alpha^\text{vir} t_\alpha^i f_i^\alpha,
\label{DeltaEHFCABS}
\end{equation}
where $i$ runs over active HF orbitals and $\alpha$ runs over all virtual orbitals (obtained in the normal basis set $\B$ and from the CABS). In Eq.~(\ref{DeltaEHFCABS}), $f_i^\alpha$ are Fock matrix elements and $t_\alpha^i$ are single-excitation coefficients found by solving the first-order perturbation equations
\begin{equation}
f_\alpha^i = \sum_j^\text{act} t_\alpha^j f_j^i - \sum_\beta^\text{vir} f_\alpha^\beta t_\beta^i.
\end{equation}
The correction is often referred to as the CABS single-excitation correction. Note that a similar correction is used in the dual basis-set approach proposed by Wolinski and Pulay~\cite{WolPul-JCP-03} for improving HF energies and by Liang and Head-Gordon~\cite{LiaHea-JPCA-04} for Kohn-Sham DFT energies.

The total basis-set corrected MP2 energy is thus
\begin{equation}
E_\text{MP2+CABS+DFT}^\B = E_\text{MP2}^\B + \Delta E_\text{HF}^{\B,\text{CABS}} + \bar{E}^\B[n_\HF^\B],
\end{equation}
and will be referred to as ``MP2+CABS+DFT''. For comparison, we will also present MP2 results only corrected by the CABS single-excitation correction, which will referred to as ``MP2+CABS''.

\section{Computational details}
\label{sec:details}

The DBBSC method with density fitting has been implemented in the \textsc{MOLPRO} software~\cite{WerKnoKniManSch-WIR-12,WerKnoManBlaDolHesKatKohKorKreMaMilMitPetPolRauSib-JCP-20,Molproshort-PROG-23}. We have performed tests on the FH51 benchmark set. The FH51 set~\cite{FriHan-JCTC-13,Fri-JCTC-15} is a set of 51 reaction energies for various organic molecules. It is included in the GMTKN55 database~\cite{GoeHanBauEhrNajGri-PCCP-17}. The FH51 set contains a large variety of molecules of different sizes (from 2 to 29 atoms). It is thus suitable to test the DBBSC method over systems of different sizes. Moreover, many systems are large enough so that the present density-fitting implementation has a large impact on the performance of the method. As regards the basis set $\B$, we use the aug-cc-pV$n$Z basis sets~\cite{KenDunHar-JCP-92} for first-row atoms and the aug-cc-pV($n$+d)Z basis sets~\cite{DunPetWil-JCP-01} for second-row atoms, which we jointly abbreviate as av$n$z, for $n=2$ (d), $3$ (t), $4$ (q), and $5$.

We perform canonical-orbital density-fitting HF~\cite{PolWerManKno-MP-04} and density-fitting MP2~\cite{WerManKno-JCP-03} calculations with the frozen-core approximation. We calculate the basis-set correction with different functionals evaluated at the active HF density, and including the CABS single-excitation correction~\cite{AdlKniWer-JCP-07,KniWer-JCP-08,KniAdlWer-JCP-09}. The basis-set correction is consistently calculated in the frozen-core approximation, corresponding to using only active orbitals in Eq.~(\ref{fBHFr}) and in the HF density used in Eq.~(\ref{EMP2+DFT}). For the $n=2,3$, for comparison, we also perform canonical-orbital density-fitting MP2-F12 (in the default 3C(F) variant with a Slater-type geminal exponent of $1$ bohr$^{-1}$)~\cite{WerAdlMan-JCP-07} calculations, implicitly including the CABS single-excitation correction. The MP2/CBS reference values are estimated as the MP2-F12 values with the av5z basis set.

For a given basis set $\B$, the density-fitting basis sets used are the corresponding $\B$/JKFIT and $\B$/MP2FIT basis sets of Weigend \textit{et al.}~\cite{Wei-PCCP-02,WeiKohHat-JCP-02} (and their extensions~\cite{KniWer-JCP-08}) for the HF and MP2 calculations, respectively. The $\B$/JKFIT basis set is also used as CABS for the CABS single-excitation correction. We have checked the density-fitting errors and found them to be insignificant. For large systems, density-fitting calculations of the basis-set correction can be more than an order of magnitude faster than non-density-fitting calculations (computation times can be found in the Supplementary Material).

\section{Results and discussion}
\label{sec:results}

\begin{figure}
\centering 
\includegraphics[scale=0.3,angle=-90]{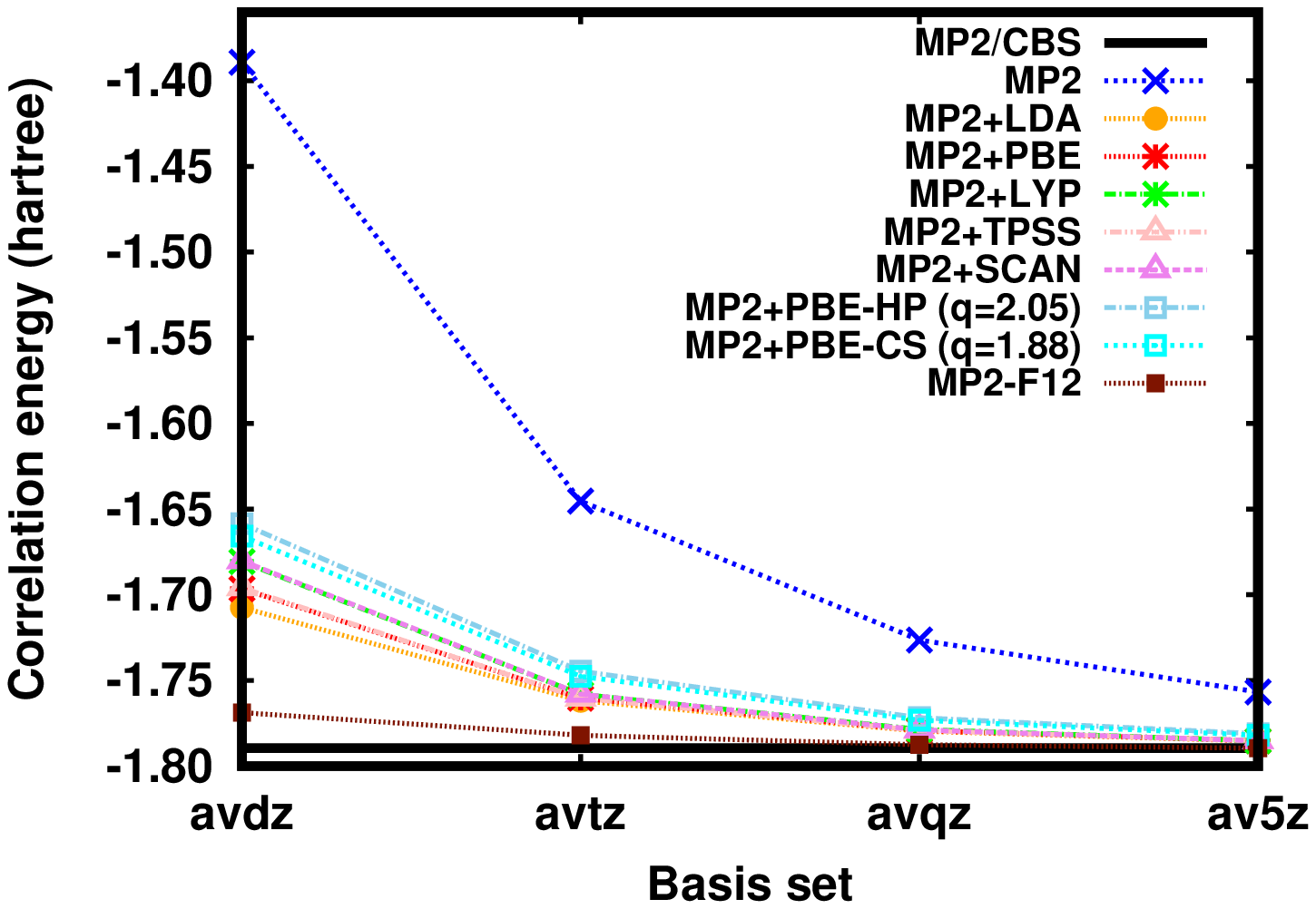}
\caption{Basis-set convergence of the ground-state MP2 correlation energy of the tetramethylpentane molecule (C$_9$H$_{20}$) with different basis-set correction density-functional approximations (evaluated at the HF density) and with MP2-F12 using av$n$z basis sets.}
\label{fig:tetramethylpentane}
\end{figure}

As a first test, we compare in Fig.~\ref{fig:tetramethylpentane} the different basis-set correction density-functional approximations for the basis-set convergence of the ground-state MP2 correlation energy of the tetramethylpentane molecule (C$_9$H$_{20}$). We see that all the density-functional approximations lead to a quite similar acceleration of the convergence of MP2 correlation energy toward its CBS limit. At the level of the correlation energy, all the proposed density-functional approximations thus provide a reasonable basis-set correction with quite a substantial acceleration of the basis-set convergence, albeit not as impressive as the one obtained with MP2-F12.

%%%%%%%%%%%%%%%%%%%%%%%%%%%%%%%%%%%%%%%%%%%%%%%%%%%%%%%%%%%%%%%%%%%%%%%%%%%%%%%%%%
%%%%%%%%%%%%%%%%%%%%%%%%%%%%%%%%%%%%%%%%%%%%%%%%%%%%%%%%%%%%%%%%%%%%%%%%%%%%%%%%%%

\begin{figure*}
\centering 
\includegraphics[scale=0.3,angle=-90]{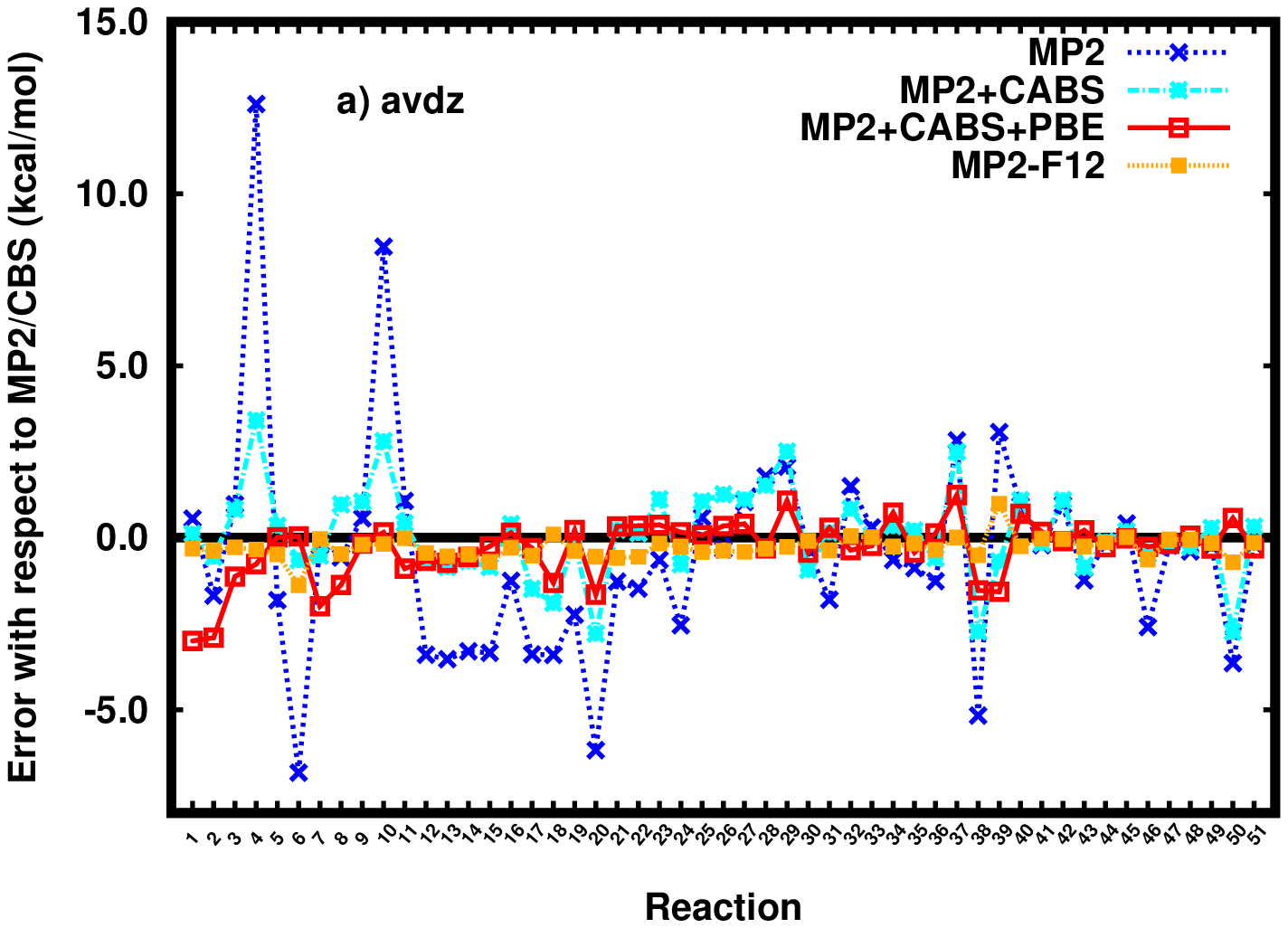}
\includegraphics[scale=0.3,angle=-90]{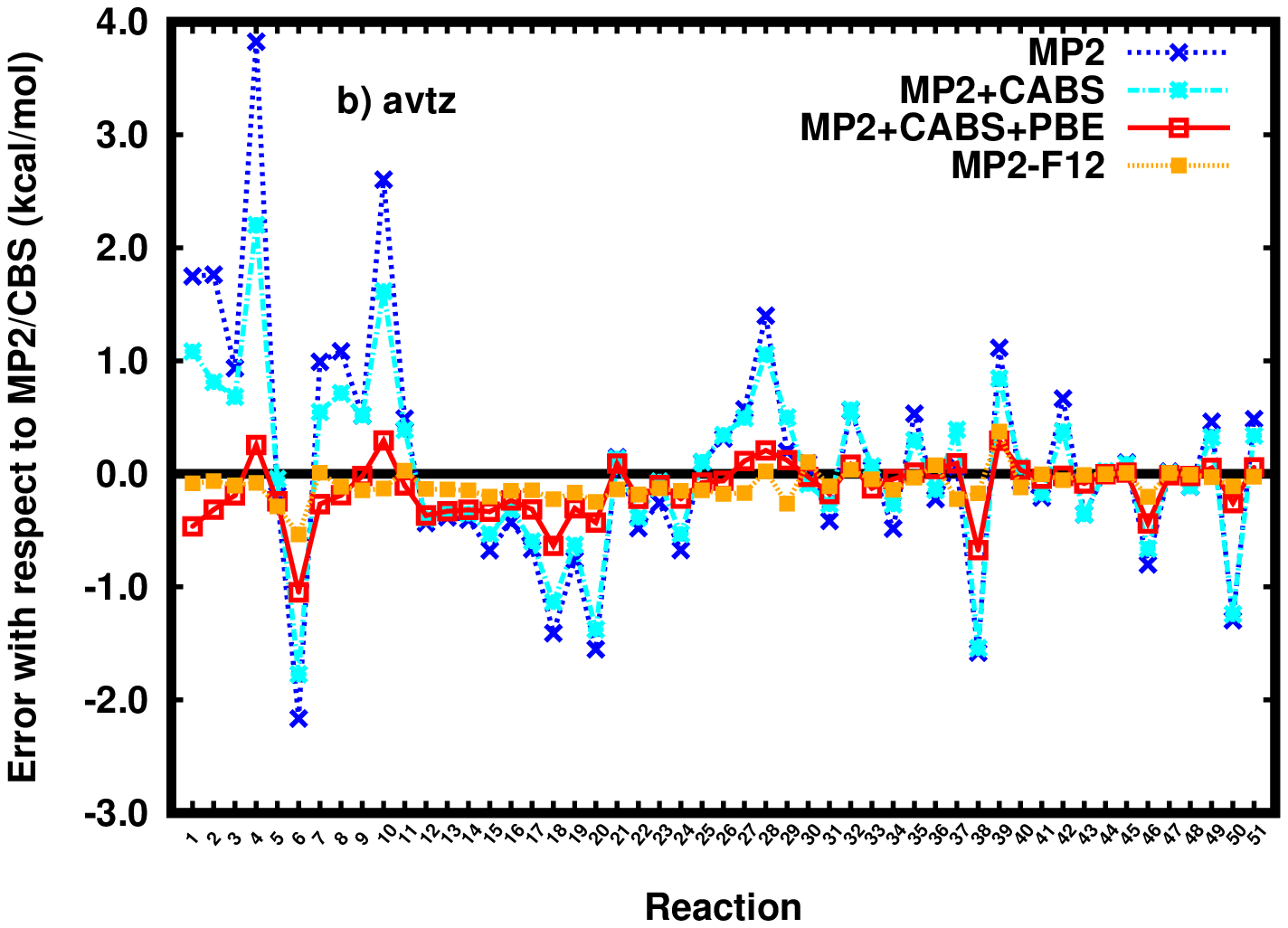}
\includegraphics[scale=0.3,angle=-90]{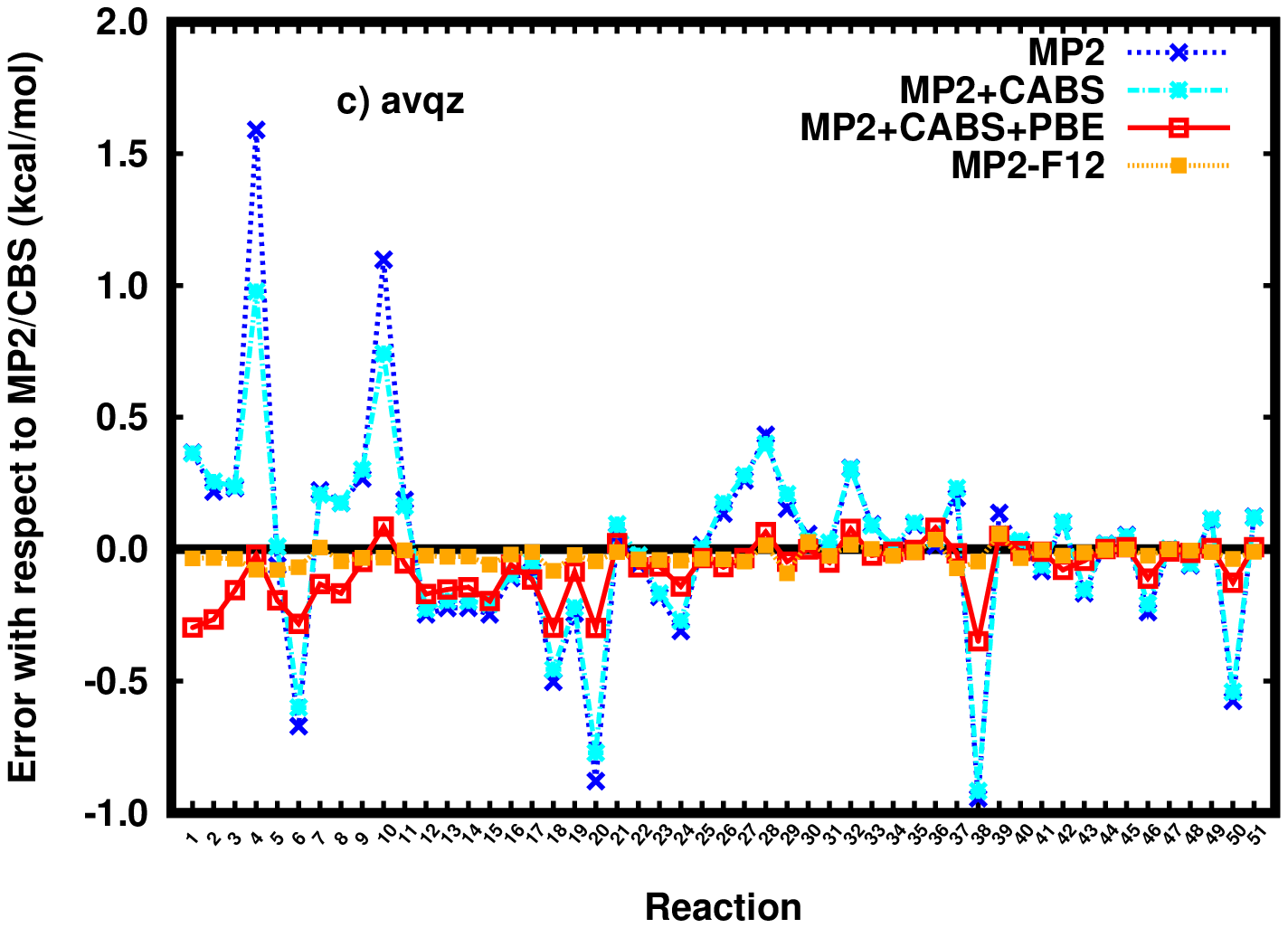}
\includegraphics[scale=0.3,angle=-90]{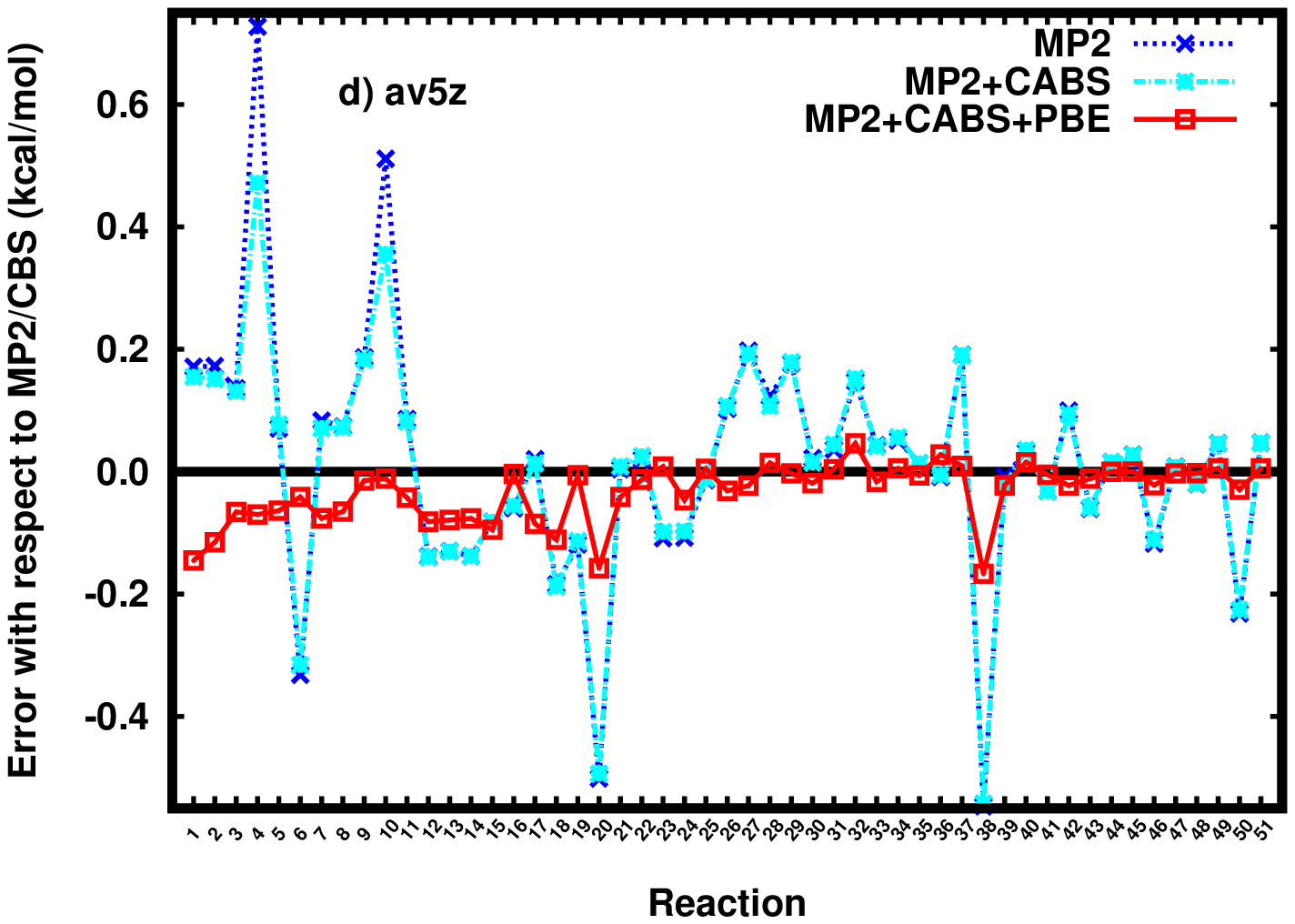}
\caption{Errors in reaction energies of the FH51 set with respect to MP2/CBS calculated with MP2, MP2+CABS, MP2+CABS+PBE, and MP2-F12 with av$n$z basis sets. The order of reactions is the one from Refs.~\onlinecite{FriHan-JCTC-13,Fri-JCTC-15}.}
\label{fig:FH51err}
\end{figure*}

The errors on the reaction energies of the FH51 set with respect to MP2/CBS calculated with MP2, MP2+CABS, MP2+CABS+PBE, and MP2-F12 are reported in Fig.~\ref{fig:FH51err}. With the avdz basis set, MP2 can have quite large basis errors for some reaction energies, up to about 13 kcal/mol. Obtaining MP2 reaction energies with all basis errors below 1 kcal/mol requires the use of the av5z basis set. The CABS single-excitation correction is crucial to reduce the largest basis errors on MP2 reaction energies obtained with the avdz basis set. Even with larger basis sets, the CABS single-excitation correction still helps to reduce the basis errors for some reaction energies. Adding the PBE-based basis-set correction further reduces the basis errors, albeit not always in a systematic way since there are a few cases where the basis error increases. It is noteworthy that the basis errors of the MP2+CABS+PBE reaction energies are all smaller than 1 kcal/mol with the avtz basis set and larger basis sets. MP2-F12 globally outperforms MP2+CABS+PBE, giving reaction energies with basis errors below about 1 kcal/mol already with the avdz basis set.

In Table~\ref{tab:FH51err}, we report the mean absolute errors (MAEs) on the reaction energies of the FH51 set with respect to MP2/CBS obtained with the methods already discussed, as well as with additional basis-set correction functionals, namely LDA, LYP, TPSS, SCAN, PBE-CS ($q=1.88$), and PBE-HP ($q=2.05$). For the methods already discussed, the mean errors are consistent with the observations made previously. For the avdz basis set, we go from a MAE of 2.08 kcal/mol for uncorrected MP2 to a MAE of 0.62 kcal/mol for MP2+CABS+PBE and a MAE of 0.33 kcal/mol for MP2-F12. For the avtz basis set, we go from a MAE of 0.72 kcal/mol for uncorrected MP2 to a MAE of 0.21 kcal/mol for MP2+CABS+PBE and a MAE of 0.13 kcal/mol for MP2-F12. For the avqz and av5z basis sets, the PBE-based basis-set correction is still effective in reducing the basis errors, as we go from MAEs of 0.25 and 0.13 kcal/mol, respectively, for uncorrected MP2 to MAEs of 0.10 and 0.04 kcal/mol, respectively, for MP2+CABS+PBE. Thus, MP2+CABS+PBE with an av$n$z basis set globally gives uncorrected MP2 reaction energies with slightly higher av$(n+1)$z quality, whereas MP2-F12 with an av$n$z basis set roughly gives uncorrected MP2 reaction energies with slightly lower av$(n+2)$z quality.

%%%%%%%%%%%%%%%%%%%%%%%%%%%%%%%%%%%%%%%%%%%%%%%%%%%%%%%%%%%%%%%%%%%%%%%%%%%%%%%%%%
%%%%%%%%%%%%%%%%%%%%%%%%%%%%%%%%%%%%%%%%%%%%%%%%%%%%%%%%%%%%%%%%%%%%%%%%%%%%%%%%%%

%%%%%%%%%%%%%%%%%%%%%%%%%%%%%%%%%%%%%%%%%%%%%%%%%%%%%%%%%%%%%%%%%%%%%%%%%%%%%%%%%%
\begin{table}
\caption{Mean absolute errors (in kcal/mol) in reaction energies of the FH51 set with respect to MP2/CBS with av$n$z basis sets.}
\begin{tabular}{lcccc}
\hline\hline
              & avdz & avtz & avqz & av5z\\
\hline
MP2           & 2.08  & 0.72 & 0.25 & 0.13\\
MP2+CABS      & 0.94  & 0.55 & 0.22 & 0.12\\
MP2+PBE       & 1.73  & 0.34 & 0.13 & 0.04\\
MP2+CABS+PBE  & 0.62  & 0.21 & 0.10 & 0.04\\
MP2+CABS+LDA  & 0.67  & 0.19 & 0.09 & 0.04\\
MP2+CABS+LYP  & 1.12  & 0.46 & 0.22 & 0.21\\
MP2+CABS+TPSS & 0.63  & 0.21 & 0.10 & 0.04\\
MP2+CABS+SCAN & 0.66  & 0.23 & 0.10 & 0.04\\
MP2+CABS+PBE-CS ($q=1.88$) & 0.62  & 0.22 & 0.09 & 0.04\\
MP2+CABS+PBE-HP ($q=2.05$) & 0.64  & 0.22 & 0.10 & 0.04\\
MP2-F12       & 0.33  & 0.13 & 0.03 & 0.00\\
\hline\hline
\end{tabular}
\label{tab:FH51err}
\end{table}
%%%%%%%%%%%%%%%%%%%%%%%%%%%%%%%%%%%%%%%%%%%%%%%%%%%%%%%%%%%%%%%%%%%%%%%%%%%%%%%%%%

With the other basis-set correction functionals tested, the MAEs are very similar, except for the LYP correlation functional which gives much larger basis errors. We have also tested optimizing the parameter $q$ in the CS and HP on-top pair density-density models in Eq.~(\ref{betan}) and the parameter $c$ in front of the on-top pair density in Eq.~(\ref{ecmdsr}), but we did not obtain significant improvements. Thus, if we set aside LYP, we find a rather small sensitivity of the method to the underlying correlation functional for calculating reaction energies. In the Supplementary Material, we report additional statistical indicators which confirm this conclusion.

Finally, as regards the computational cost of the DFT-based basis-set correction in comparison with MP2-F12, we consistently observe, for all basis sets, that MP2+CABS+PBE is approximately 10 times faster than MP2-F12 in the default 3C variant. However, we note that MP2-F12 can be made faster using the 3*A approximation~\cite{WerAdlMan-JCP-07} without losing much accuracy in most cases, and MP2+CABS+PBE is only approximately 3 to 4 times faster than this cheaper MP2-F12 variant. Of course, the relative gains in computational cost would be much less for more expensive wave-function methods such as CCSD(T).

\section{Conclusion}
\label{sec:conclusion}

We have reported an efficient density-fitting implementation of the DBBSC method in the \textsc{MOLPRO} software using different basis-set correction density-functional approximations and including the CABS single-excitation correction. We have tested the method on the FH51 benchmark set of reaction energies at the MP2 level and provided a comparison with the explicitly correlated MP2-F12 method.

For the smallest basis sets, the CABS single-excitation correction provides an important correction on reaction energies which is not included in the basis-set correction density-functional approximations. The basis-set corrected reaction energies are quite insensitive to the choice of the basis-set correction density-functional approximation, with the notable exception of the LYP functional which gives much worse results. This point should be further analyzed in the future. Overall, the basis-set corrected MP2 reaction energies calculated with a $n$-zeta basis set are of slightly higher quality than uncorrected MP2 reaction energies calculated with $(n+1)$-zeta quality. However, the explicitly correlated MP2-F12 method is consistently more accurate, with reaction energies calculated with a $n$-zeta basis set being of slightly lower quality than uncorrected MP2 reaction energies calculated with $(n+2)$-zeta quality. We believe that the DBBSC method is still valuable for accelerating the basis convergence of MP2 due to the fact that it has a lower computational cost than MP2-F12.

Finally, let us mention that the present implementation of the DBBSC method can be applied to any other wave-function methods, such as CCSD(T), with expected similar gains in accuracy. After completion of the present work, we became aware of a very similar independent work that has just been published~\cite{MesKal-JCTC-23}.

\begin{acknowledgments}
It is a pleasure to dedicate the present paper to Carlo Adamo on the occasion of his 60th birthday.
\end{acknowledgments}

\section*{Supplementary Material}
See the Supplementary Material for computation times and further statistical measures.

\section*{Conflict of interest}
None of the authors have a conflict of interest to disclose.

% BIBLIOGRAPHY---------------------------------------------
%\bibliographystyle{jchemphys}
%\bibliography{biblio}

\end{document}